
\input epsf
\input harvmac

\def\eg{{\it e.g.}}
\def\ie{{\it i.e.}}

\def\frac#1#2{{{#1}\over {#2}}}
\def\half{\hbox{${1\over 2}$}}

\def\smallfrac#1#2{\hbox{${{#1}\over {#2}}$}}

\def\GeV{{\rm GeV}}

\def\MS{\hbox{$\overline{\rm MS}$}}

\catcode`@=11 
\def\slash#1{\mathord{\mathpalette\c@ncel#1}}
 \def\c@ncel#1#2{\ooalign{$\hfil#1\mkern1mu/\hfil$\crcr$#1#2$}}
\def\lsim{\mathrel{\mathpalette\@versim<}}
\def\gsim{\mathrel{\mathpalette\@versim>}}
 \def\@versim#1#2{\lower0.2ex\vbox{\baselineskip\z@skip\lineskip\z@skip
       \lineskiplimit\z@\ialign{$\m@th#1\hfil##$\crcr#2\crcr\sim\crcr}}}
\catcode`@=12 
\def\twiddles#1{\mathrel{\mathop{\sim}\limits_
                        {\scriptscriptstyle {#1\rightarrow \infty }}}}
\def\PR{{\it Phys.~Rev.~}}

\def\NP{{\it Nucl.~Phys.~}}

\def\PL{{\it Phys.~Lett.~}}

\def\ZP{{\it Zeit.~Phys.~}}

\def\JHEP{{\it Jour.~High~Energy~Phys.~}}
\def\vol#1{{\bf #1}}\def\vyp#1#2#3{\vol{#1} (#2) #3}

\def\trge{transverse renormalization group equation}
\def\lrge{longitudinal renormalization group equation}

\def\as{a}
\def\bas{a}

\noblackbox
\pageno=0\nopagenumbers\tolerance=10000\hfuzz=5pt
\baselineskip 12pt
\line{\hfill {\tt hep-ph/9906222}}
\line{\hfill Edinburgh 99/5}
\line{\hfill RM3-TH/99-1}
\line{\hfill DFTT 27/99}
\vskip 12pt
\centerline{\bf  The
Small $x$ Behaviour of Altarelli-Parisi Splitting
Functions}
\vskip 18pt\centerline{Richard D. Ball\footnote{$^*$}{\footnotefont
Royal Society University Research Fellow}}
\vskip 6pt
\centerline{\it Department of Physics and Astronomy}
\centerline{\it University of Edinburgh, EH9 3JZ, Scotland}
\vskip 12pt
\centerline {Stefano Forte\footnote{$^\dagger$}{On leave from INFN, 
Sezione di Torino, Italy}}
\vskip 6pt
\centerline {\it I.N.F.N., Sezione di Roma III}
\centerline {\it Via della Vasca Navale 84, I-00146 Rome, Italy}
\vskip 50pt
\centerline{\bf Abstract}
{\narrower\baselineskip 10pt
\medskip\noindent
We extract the small $x$ asymptotic behaviour of the 
Altarelli-Parisi splitting functions from their expansion 
in leading logarithms of $1/x$.
We show in particular that the nominally next-to-leading correction 
extracted from the Fadin-Lipatov kernel is enhanced asymptotically 
by an extra $\ln {1\over x}$ over the leading order. 
We discuss the origin of this problem, 
its dependence on the choice of factorization scheme, and its all-order
generalization. We derive necessary conditions which must be 
fulfilled in order to obtain a well behaved perturbative expansion, 
and show that they may be satisfied by a suitable reorganization 
of the original series.
\smallskip}
\vfill
\line{June 1999\hfill}
\eject \footline={\hss\tenrm\folio\hss}


\lref\rev{See \eg\ R.D.~Ball and A.~DeRoeck, {\tt hep-ph/9609309}\semi
S.~Forte, {\tt hep-ph/9812382} and ref. therein.} 
\lref\DGPTWZ{A.~De~R\'ujula et al., \PR\vyp{10}{1974}{1649}.}
\lref\das{R.D.~Ball and S.~Forte,
\PL\vyp{B335}{1994}{77}.}
\lref\summ{R.D.~Ball and S.~Forte, \PL\vyp{B351}{1995}{313}\semi 
R.K.~Ellis, F.~Hautmann and B.~R.~Webber, \PL\vyp{B348}{1995}{582}.}
\lref\fits{R.D.~Ball and S.~Forte, {\tt hep-ph/9607291}\semi see also
I.~Bojak and M.~Ernst, \NP\vyp{B508}{1997}{731}.}
\lref\fl{V.~S.~Fadin and L.~N.~Lipatov, \PL\vyp{B429}{1998}{127}.}
\lref\flx{V.~S.~Fadin et al, \PL\vyp{B359}{1995}{181};
\PL\vyp{B387}{1996}{593};
\NP\vyp{B406}{1993}{259}; \PR\vyp{D50}{1994}{5893}; 
\PL\vyp{B389}{1996}{737};
\NP\vyp{B477}{1996}{767}; \PL\vyp{B415}{1997}{97};
\PL\vyp{B422}{1998}{287}.} 
\lref\flxx{V.~del~Duca, \PR\vyp{D54}{1996}{989};
\PR\vyp{D54}{1996}{4474} \semi V.~del~Duca and C.~R.~Schmidt, 
\PR\vyp{D57}{1998}{4069} \semi 
Z.~Bern, V.~del~Duca and C.~R.~Schmidt, \PL\vyp{B445}{1998}{168}
\semi  G.~Camici and M.~Ciafaloni, 
\PL\vyp{B412}{1997}{396}; \PL\vyp{B430}{1998}{349}.}
\lref\brus{R.~D.~Ball  and S.~Forte, {\tt hep-ph/9805315}.}
\lref\blum{J. Bl\"umlein et al., {\tt hep-ph/9806368}.}
\lref\ross {D.~A.~Ross, \PL\vyp{B431}{1998}{161}.}
\lref\levin{E.~Levin, {\tt hep-ph/9806228}.}
\lref\muel{Y.~V.~Kovchegov and A.~H.~Mueller,
\PL\vyp{B439}{1998}{428}.}
\lref\ABB{N.~Armesto, J.~Bartels and M.A.~Braun, 
\PL\vyp{B442}{1998}{459}.}
\lref\afp{R.~D.~Ball and S.~Forte, {\it Phys. Lett.} {\bf
B405}, 317 (1997).}
\lref\oldls{ T.~Jaroszewicz, \PL\vyp{B116}{1982}{291}\semi
S.~Catani, F.~Fiorani and G.~Marchesini, \PL\vyp{B336}{1990}{18}
\semi S.~Catani et al.,  \NP\vyp{B361}{1991}{645}.}
\lref\bfkl{For a review, see  
V.~del~Duca, {\tt hep-ph/9503226} and ref. therein.}
\lref\ciaf{G.~Camici and M.~Ciafaloni, 
\NP\vyp{B496}{1997}{305}.}
\lref\CH{S.~Catani and F.~Hautmann, \PL\vyp{B315}{1993}{157}; 
\NP\vyp{B427}{1994}{475}.}
\lref\bfapp{R.D.~Ball and S.~Forte, unpublished.}
\lref\fad{V.S.~Fadin, {\tt hep-ph/9807527}}
\lref\ciafqz{M.~Ciafaloni, \PL\vyp{B356}{1995}{74}.}
\lref\sdis{S.~Catani, \ZP\vyp{C70}{1996}{263}.}
\lref\phys {S.~Catani,  {\it Z. Phys.} {\bf C75}, 665 (1997).}
\lref\mom{R.~D.~Ball and S.~Forte, {\it Phys. Lett.} {\bf
B359}, 362 (1995).}
\lref\blm{S.J.~Brodsky et al, {\tt hep-ph/9901229}.}
\lref\rst{R.S.~Thorne, {\tt hep-ph/9901331}.} 
\lref\salami{G.~Salam, \JHEP\vyp{9807}{1998}{19}\semi
M.~Ciafaloni and D.~Colferai, {\tt hep-ph/9812366}\semi
C.R.~Schmidt, {\tt hep-ph/9901397}\semi
J.R.~Forshaw, D.A.~Ross and A.~Sabio~Vera, {\tt hep-ph/9903390}.}
 

The inclusive structure function $F_2(x,Q^2)$
has been determined with extraordinary accuracy down to very small
$x$ by recent experiments at the HERA collider~\rev. For 
$Q^2\gsim 1\GeV^2$ the $x$ and $Q^2$ dependence of the data is
in complete agreement with that predicted by the next-to-leading 
order (NLO) Altarelli-Parisi evolution equations. 
At small $x$ and large $Q^2$, the evolution equations are 
dominated by the small $x$ singularities of the Altarelli-Parisi 
splitting functions~\DGPTWZ, and retaining only the 
singularities in the LO and NLO splitting functions
yields an excellent approximation to the full solution in the HERA
region~\das. 

As we go to higher orders in $\alpha_s$ the splitting functions 
become more and more singular,
and these higher order singularities might be expected to become dominant
at small enough $x$. It thus appears reasonable to try to improve
the description of small $x$ evolution by supplementing the
usual leading-order splitting functions with contributions
which sum all leading logs of $x$ (LLx) to all orders in $\alpha_s$,
\ie\ all terms of the form $(\alpha_s\log x)^n$. 
Likewise, the NLO splitting functions can be supplemented by 
a summation of next-to-leading log $x$ contributions (\ie\ all terms
of the form $\alpha_s(\alpha_s\log x)^n$, and so 
on (the `double-leading expansion'~\summ).

However, as is by now well known~\fits, such attempts are 
unsuccessful: essentially, the data in the HERA region
are so accurately described by plain NLO evolution (and the
small $x$ approximation to it) that any further ``improvement'' 
would spoil this agreement unless its effects were extremely small. 
Furthermore, the recent determination~\refs{\fl,\flx,\flxx} of 
the subleading corrections
to LLx evolution has shown that NLLx contributions are 
extremely large, and in fact grow faster as $ x\to0$ than their LLx
counterparts~\refs{\brus,\blum}. Hence, contrary to naive expectations, the 
double leading expansion does not appear to be stable at small $x$.
A deeper understanding of the all-order behaviour of splitting
functions at small $x$ is required.

In this letter we will determine the small $x$ behaviour of the
Altarelli-Parisi splitting functions order by order in LLx, NLLx,\dots.
All our discussion will be
based on a formal perturbative treatment of the small $x$ expansion: 
at NLLx we retain only
the contributions to small $x$ evolution calculated in~\fl,
systematically discarding any terms which are formally NNLLx.
This approach allows us to isolate the reason for the asymptotic 
breakdown of the small $x$ expansion.  
This turns out to be unrelated to various problems discussed 
elsewhere, such as the unphysical behaviour of the solutions 
found in \refs{\ross,\levin} and the running 
coupling resummation effects discussed in \refs{\muel,\ABB}.
We find that the instability observed at NLLx 
is not some peculiar feature of the calculation of ~ref.\fl, but is
completely generic, probably persisting to all orders. Formally 
subleading contributions to the splitting function are not suppressed 
by powers of $\alpha_s$, but rather grow faster and faster at 
small $x$. We derive conditions on the small $x$ evolution
kernel which are necessary for stable evolution, and show that they
may be satisfied by a suitable reorganization of the perturbative 
expansion. Such a resummation introduces an a priori undetermined 
parameter which describes the resummed all-order  
small-$x$ behaviour of the structure functions.

The small $x$ behaviour of splitting functions $P(x)$ is most easily studied
by considering their Mellin transforms, the anomalous dimensions
$\gamma(N)\equiv\int_0^1\!dx\,x^{N} P(x)$. 
The leading small $x$ behaviour is then found by expanding 
$\gamma(N)$ about its rightmost
singularity in the complex $N$ plane, which in the singlet sector is
located at $N=0$ (and at $N=-1$ in the nonsinglet sector, which is thus
suppressed by a power of $x$ and hence asymptotically negligible).
The anomalous dimensions in the singlet sector are given by a
two-by-two matrix; however only one of the two eigenvalues of the
matrix is singular at $N=0$ at leading order (and to all orders in
appropriate factorization schemes). It is thus sufficient to concentrate
on this leading eigenvalue and its associated
eigenvector $G_N(Q^2)$, the Mellin transform of the distribution
function $G(x,Q^2)$, which satisfies the evolution equation
\eqn\trge{{d\over dt}G_N(Q^2)=\gamma(N;\as)G_N(Q^2),}
where $t\equiv\ln(Q^2/\Lambda^2)$ and for future convenience we write
$a(t)\equiv\alpha_s(t)/2\pi$.

The general structure of the anomalous dimension in the small $x$
expansion is 
\eqnn\lsexp\eqnn\expdef
$$\eqalignno{
&\gamma(N;\as)=\gamma_0(\bas/N)+
\as\gamma_1(\bas/N)+\dots
&\lsexp\cr &\qquad\gamma_k\left(
{\bas\over N}\right)\equiv \sum_{n=1}^\infty A^{(n)}_k\left(
{8N_c\ln 2}{\bas\over N}\right)^n.&\expdef\cr}$$
where the coefficents $A^{(n)}_k$ have been normalized such that
the radius of convergence of the series is one.
The associated splitting functions
$P_k$  are immediately obtained by inverse Mellin transformation
of $\gamma_k$:
\eqnn\lsexpsf\eqnn\expdefsf
$$\eqalignno{&P(x;\as)= {a\over x}
\left[P_0(\Xi)+\as P_1(\Xi)+\dots
\right]&\lsexpsf\cr
&\qquad P_k(\Xi)
=\sum_{n=1}^\infty A^{(n)}_k{\Xi^{n-1}
\over (n-1)!};&\expdefsf\cr}$$
where $\xi=\ln{1\over x}$ and we define 
\eqn\xidef{\Xi\equiv (2C_A)(4\ln 2)\smallfrac{\alpha_s}{2\pi}\xi
= 8N_c\ln 2\, \bas \xi.}

Subsequent terms $\gamma_k$, $P_k$  in the expansions
\expdef,\expdefsf\ of the
anomalous dimension and splitting function 
sum the leading, next-to-leading,\dots\ logarithms of
$1\over x$. They can therefore be determined~\refs{\oldls,\afp} from 
knowledge of the respective leading, next-to-leading,\dots\ QCD 
high-energy asymptotics, as given by leading log $x$ evolution, 
which is in turn controlled by an equation of the form
\eqn\lrge{{d\over d\xi}G_M(x)=\bas\chi(M;\as)G_M(x),}
where the Mellin variable M is defined by
\eqn\mmeldef{ G_M(x)\equiv \int_0^\infty {d Q^2\over Q^2} 
\left({Q^2\over\Lambda^2}\right)^{-M} G(x,Q^2)}
and the anomalous dimension $\chi(M;\as)$ admits a perturbative 
expansion
\eqn\chiexp{
\chi(M;\as)=\chi_0(M)+\as \chi_1(M)+\dots\qquad .}
The leading-order term $\chi_0(M)=2N_c[2\psi(1)-\psi(M)-\psi(1-M)]$ 
is well known~\bfkl, while the next-to-leading term $\chi_1$
has been determined only recently~\fl.\foot{
Note that here we have adopted a different set of normalization
conventions to those used in ref.~\fl, where the right hand side of 
eq.~\chiexp\ is written as $2N_c(\chi+ {N_c\alpha_s\over 4\pi}
\tilde\delta)$: our $\chi_0$ is the same as their $2N_c\chi$, while 
our $\chi_1$ is the same as their $N_c^2\tilde\delta$.}
Note that since structure functions scale in the $Q^2\to\infty$
limit and drop linearly with $Q^2$ as $Q^2\to0$ the leading-twist
physical region corresponds to $0<M<1$.   

The $k$-th order small $x$ anomalous
dimensions $\gamma_k\left({\as\over N}\right)$ eq.~\lsexp\ can be
determined from $\chi_0,\dots,\chi_k$ eq.~\chiexp\ 
by matching the solutions to the respective evolution
equations. Assume for the moment that the coupling is fixed. 
Solving eq.~\lrge\ by Mellin transform with respect to $N$,
we find
\eqn\nmsol{G_{NM}={G^0(M) \over N- \bas \chi(M;\as)},}
where $G^0(M)$ is the boundary condition at $\xi=0$.  
In order to compare to eq.~\trge, invert the $M$-Mellin at
large $Q^2$:
\eqn\tnsol{G_{N}(Q^2)={G^0[M_p(N;\as)]
\over (-\bas \chi'[M_p(N;\as)])}
e^{M_p(N;\as) t},}
where $M_p(N;\as)$ is the position of the 
rightmost pole of $G_{NM}$ in the $M$-plane in the physical region $0<M<1$. 
It follows that the solutions to the evolution equations
\trge\ and~\lrge\ coincide only if $\gamma=M_p$, \ie\ 
if their anomalous dimensions satisfy the `duality' relation~\afp\
\eqn\dual{\chi[\gamma(N;\as);\as]={N\over\bas}.}

Expansion of eq.~\dual\ in powers of $\as$ keeping $\bas/N$
fixed gives a set of relations
which determine $\gamma_k$ perturbatively order by order in terms of 
the perturbative expansion of $\chi$:
\eqnn\expduallo\eqnn\expdualnlo
\eqnn\expdualnnlo
$$\eqalignno{&\chi_0[\gamma_0({\bas/N})]=
{N\over\bas}&\expduallo\cr
&\gamma_1\left({\bas/N}\right)=-{\chi_1[\gamma_0(\bas/N)]
\over{\chi_0}'[\gamma_0(\bas/N)]}&\expdualnlo\cr
&{\gamma_2({\bas/N})}=-{\chi_2 ({\chi_0}')^2
-\chi_{1}{\chi_{1}}'{\chi_{0}}'+{1\over2}(\chi_{1})^2 {\chi_{0}}''
\over({\chi_{0}}')^3},&\expdualnnlo\cr}$$
and so forth, where the prime indicates differentiation  
with respect to $M$, and in the last equation 
all $\chi$-functions have argument $\gamma_0(\bas/N)$. 
Eq.~\expduallo\ should be viewed as an implicit equation for
$\gamma_0$, with $\chi_0(M)$ evaluated in the physical region $0<M<1$.

The derivation presented so far
holds at fixed coupling $\alpha_s$. If the coupling 
runs with $Q^2$ in both equations, it is sufficient to include
the running up to order $\alpha_s^k$ when computing the anomalous
dimensions to $k$-th order in the LLx expansion. The contributions 
to the kernel on the right hand side of eq.\lrge\ are then found
by replacing  $\alpha_s$ by differential operators: 
for example, to NLLx it is sufficient to let  
\eqn\alfrun{
\as \to  \as\left(1 + b_0 \as \frac{d}{dM}\right),}
where $b_0 = 
\half(\smallfrac{11}{3}N_c-\smallfrac{2}{3}n_f)$. Solving \lrge\ as before,
taking care to treat all NLLx terms (including the derivative term) 
as perturbations, now gives
\eqn\nmsolrun{
G_{NM}=\left[{1 \over 1- {\bas\over N} \chi_0} + 
 \as  \frac{\bas}{N}
\Big(\frac{ \chi_1 
+b_0\chi_0(\ln\chi_0 G_0)'}{(1- {\bas\over N}\chi_0)^2}
+\frac{b_0\chi_0 \frac{\bas}{N}\chi'_0}
{(1- {\bas\over N}\chi_0)^3}\Big)+\ldots  
\right]{G^0(M)\over N}.
}
Inverting the $M$-Mellin, and again comparing with the solution of eq.~\trge\
with the running coupling expanded at NLLx, we see that the solutions match
if in place of \expdualnlo\ we now have
\eqnn\nlochi\eqnn\nloqz
$$\eqalignno{\gamma_1(\bas/N)
&=-\frac{1}{{\chi_0}'(\gamma_0)} \left(\chi_1(\gamma_0)
+\half b_0\frac{\chi_0(\gamma_0) \chi_0''(\gamma_0)}
{\chi_0'(\gamma_0)}\right)&\nlochi\cr
&=-\frac{\chi_1(\gamma_0)}{{\chi_0}'(\gamma_0)}
- \frac{1}{\as}\frac{d}{dt} \ln \sqrt{-\chi_0'( \gamma_0)},&\nloqz\cr }$$
since to leading order $d\as/dt=-b_0\as^2$, while 
$\bas\chi_0(\gamma_0)=N$ so $\partial\gamma_0/\partial\ln\as =
-\chi_0/\chi_0'$. The expression \nloqz\ was obtained in \ciaf\ by 
solving eq.\lrge\ with running coupling exactly, and then 
inverting the Mellin by a saddle point argument consistently 
at NLLx.

When the coupling runs there is however a further ambiguity related to
the choice of factorization scheme. Under a LLx change in the 
normalization of $G_M(\xi)$, \ie\  $G_M(\xi) \to u(M)G_M(\xi)$, 
the NLLx kernel changes according to 
\eqn\schchi{\chi_1(M)\to \chi_1(M) - b_0\chi_0(M) 
\frac{d}{dM} \ln u(M).}
{}From eq.\nloqz\ the anomalous dimension changes as
\eqn\schgam{\gamma\to\gamma 
+ \frac{d}{dt} \ln u(\gamma_0).}
This is the same as the NLLx shift in the anomalous dimension induced by
the LLx scheme change  $G_N(t) \to u(\bas(t)/N)G_N(t)$,
where $u(\bas/N)=1+\sum_1^\infty u_n(\bas/N)^n$ \mom. 
Note that a NLLx scheme change only affects the anomalous dimension at
NNLLx: the mismatch is due to the coupling running with $Q^2$ while
the logarithms are ordered in $x$. It follows that knowledge
of the leading order coefficient function \CH\ is sufficient for 
a consistent calculation at NLLx.

In order to determine the NLLx anomalous dimension $\gamma_1$ in 
\MS\ factorization\foot{
Note that the large eigenvalue of evolution at small $x$ is invariant
under the usual (NLLQ) scheme changes: it follows that for our
purposes the \MS\ scheme is the same as the DIS scheme, and indeed
we will use these two notations interchangeably.} from the 
Fadin-Lipatov kernel $\chi_1^{\rm FL}$
(eqns.(14)\& (22) of ref.\fl) two further adjustments are required.  
First, the kernel $\chi_1$ for the small $x$ evolution of the
distribution $G_M(\xi)\equiv M\inv g_M(\xi)$ is related to 
that for the unintegrated distribution $g_M(\xi)$ employed in \fl\ by 
$\chi_1(M) = \chi_1^{\rm FL}(M) + b_0\chi_0(M)/M$. This may  
be thought of as a LLx scheme change with $u(M)=M$. Second, 
the correct expression for $\gamma_1$ with \MS\ factorization 
requires a further scheme change with
$u(M)=R(M)$, where $R(\gamma_0(\as/N)))\equiv R_N(\alpha_s)$ 
is calculated in ref.\CH.\foot{To see this compare
the result for $\gamma_{qg}$ in \MS\ obtained in \CH\ with that 
derived in \ciafqz: their ratio must be $u$. Alternatively, 
the explicit derivation given in Appendix B of \CH\ of the 
anomalous dimension from the dimensionally regularized kernel 
may be extended to NLLx: this gives the same result \bfapp.}
The \MS\ NLLx anomalous dimension is thus
\eqnn\gammsb\eqnn\chimsb
$$\eqalignno{\gamma_1&(\bas/N)
=-\frac{\chi_1^{FL}(\gamma_0)}{{\chi_0}'(\gamma_0)}
+ \frac{1}{a}\frac{d}{dt} 
\ln R(\gamma_0)\gamma_0\sqrt{-\chi_0'( \gamma_0)}&\gammsb\cr 
&=-\frac{1}{{\chi_0}'(\gamma_0)} \left(\chi_1^{FL}(\gamma_0)
+b_0 N_c
((2\psi'(1)-\psi'(\gamma_0)-\psi'(1-\gamma_0))
+\smallfrac{1}{4N_c^2}\chi_0(\gamma_0)^2)\right),&\chimsb\cr}$$
where in the second line we used eq.(B.18) of \CH\ for 
$\partial\ln R/\partial\gamma_0$. 

\topinsert
\vskip-2.5truecm
\vbox{
\hfil\epsfxsize=8.5truecm\epsfbox{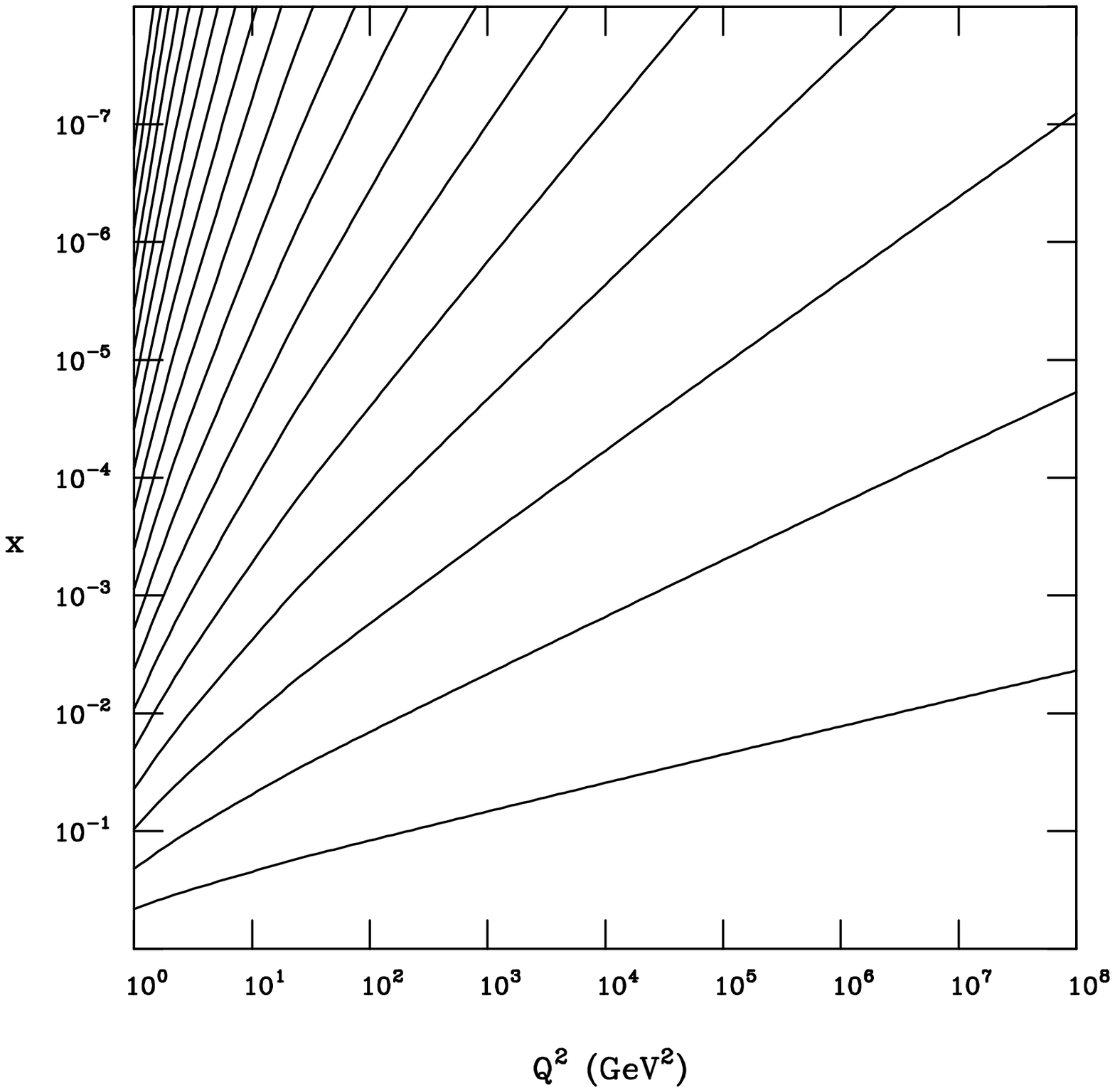}\hfil
\vskip-1.0truecm
\hskip4truecm\hbox{\vbox{\footnotefont\baselineskip6pt\narrower\noindent
Figure 1: Contours of constant $\Xi=1,2,\ldots,20$ (from bottom to top)
in the $x-Q^2$ plane.
}}\hskip4truecm}
\medskip
\endinsert

It is clear from eqns.\gammsb\ \& \chimsb\ that at NLLx the 
effect of the running of $\alpha_s$, eq.\nloqz, can be 
entirely absorbed into a shift of the NLLx anomalous
dimension $\gamma_1$ equivalent to a choice of factorization scheme. It 
turns out that this property persits to higher orders in the small $x$
expansion. We may thus view running coupling effects as a
contribution to $\chi$: henceforth we assume that $\chi$ incorporates the 
running coupling effects at the appropriate order in the 
small $x$ expansion, so that the anomalous dimension $\gamma$ is given
by the duality relation \dual.

As is well known, the BFKL kernel $\chi_{0}(M)$ is symmetric about
$M=\half$, where it has a minimum, with $\chi_0(\half)=8N_c\ln 2$. It
follows that $\gamma_{0}$, viewed as a series of powers of 
$8N_c\ln 2{\bas\over  N}$ (eq.\expdef), has unit radius of convergence. All 
higher order $\gamma_k$ will have the same radius of convergence 
in any factorization 
scheme in which all $\chi_k(M)$ are free of singularities for 
$0<M<{1\over2}$. The fact that the series for the anomalous dimension
has finite radius of convergence is important because it shows 
that the corresponding splitting function has infinite radius 
of convergence as a power series in $\Xi$, 
and can thus be used down to arbitrarily small $x$, provided a large
enough number of terms is included \summ. In the asymptotic region where 
${A^{(n+1)}_{k}/A^{(n)}_{k}}\sim 1$ the number of terms which must
be included is of order $k_c(x,Q^2)$, where $k_c$ is the
solution of the implicit equation $\Xi^{k_c}/k_c!\sim 1$, whence
$k_c \sim 2.7 \Xi$ for $\Xi\gsim 1$. It is apparent that already in the 
HERA region, where $\Xi$ can be as large as ten (see fig.~1) a large 
(though finite) number of terms should be included.\foot{It is interesting 
to contrast this with the familiar case of large $x$ 
(Sudakov) resummation, where
the series of contributions to be resummed in the anomalous dimension
is divergent and can only be treated by Borel resummation: for
meaningful results an infinite number of terms must be included 
whenever $\alpha_s \ln(1-x)\sim 1$.}

\topinsert
\vskip-2.5truecm
\vbox{
\hfil\epsfxsize=8.5truecm\epsfbox{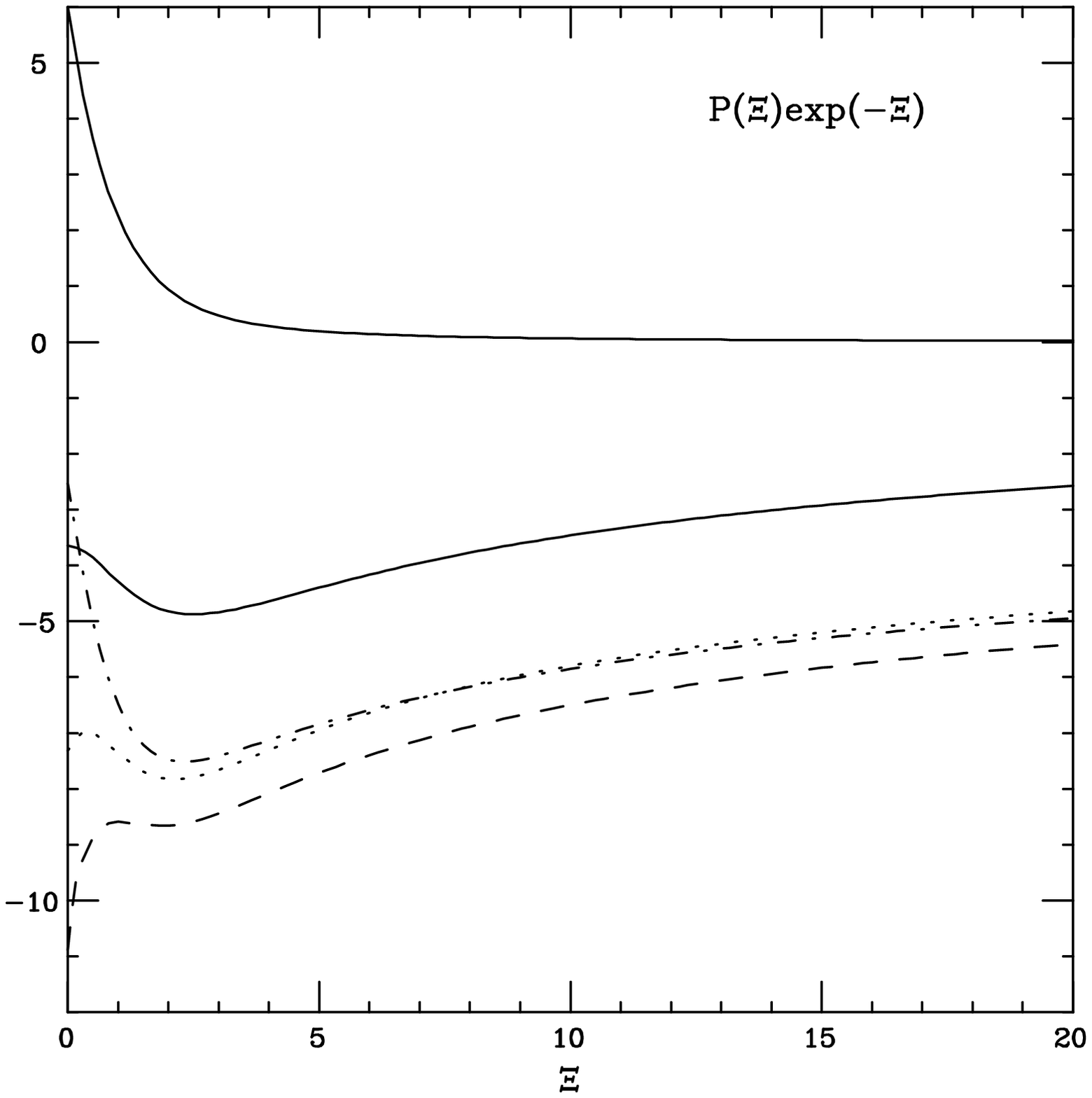}
\vskip-2.0truecm
\bigskip
\hfil\hbox{\vbox{\footnotefont\baselineskip6pt\narrower\noindent
Figure 2: The LLx splitting function $P_0(\Xi)$ (solid, positive) 
and the NLLx splitting function $\smallfrac{1}{2\pi}P_1(\Xi)$ 
(negative), computed numerically  
(using eqns. \expdefsf,\expduallo,\nloqz,\chimsb\ and 
$\chi_1^{\rm FL}$ from ref.\fl), and renormalised by a factor
$e^{-\Xi}$, in various factorization 
schemes: DIS (or \MS, see footnote 2) (solid), 
Q$_0$-DIS \ciafqz\ (dotted), SDIS \sdis\ (dashed) and 
GDIS \refs{\phys,\mom} (dot-dashed).
}}\hfil}
\medskip
\endinsert

\topinsert
\vskip-2.5truecm
\vbox{
\hfil\epsfxsize=8.5truecm\epsfbox{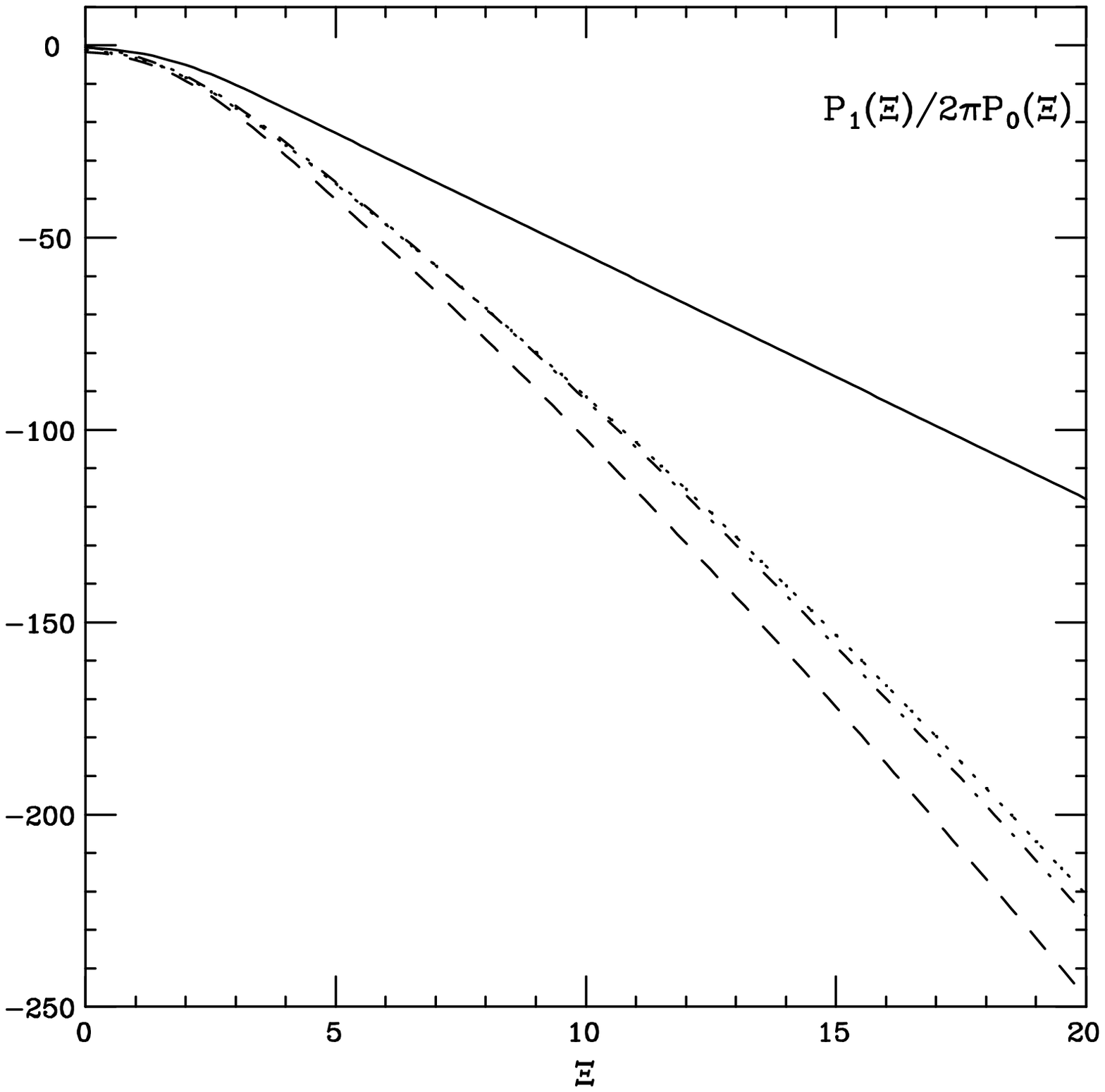}
\vskip-2.0truecm
\bigskip
\hskip4truecm\hbox{\vbox{\footnotefont\baselineskip6pt\narrower\noindent
Figure 3: The ratio ${1\over2\pi}P_1(\Xi)/P_0(\Xi)$ of NLLx and LLx splitting
functions, computed as in fig.~2, in the same factorization schemes .
}}\hskip4truecm}
\medskip
\endinsert

Since each contribution $P_k$ to the small $x$ expansion~\lsexpsf\ 
of the splitting function sums up leading logs
of $x$ to all orders, each subsequent order in the expansion 
appears to be of order $\alpha_s$ compared to the previous 
order. It was thus natural to conjecture \refs{\oldls,\CH,\summ} 
that evolution at small $x$ can be accurately described by 
truncating the expansion 
to finite order, provided  only that $\alpha_s$ is small enough
(\ie\  $Q^2$ is large enough). As discussed in the introduction,
however, this conjecture does not seem in agreement with
phenomenology~\fits, nor with the numerical comparison of the NLLx
splitting function $P_1$ to the LLx $P_0$~\refs{\brus,\blum}:
as may be seen from Fig.~2, $P_1$ is large and negative, and
furthermore the size of the ratio of $P_1$ to $P_0$ 
(see fig.~3) increases rapidly at small $x$, \ie\  as $\Xi$ increases. 
It is apparent from the plot that the ratio  ${1\over2\pi}P_1/ P_0$ 
is much greater than
$1/\alpha_s(Q^2)\lsim 10$ throughout most of the range of 
$\Xi$ relevant for HERA and the LHC. This in itself explains
the failure of the phenomenology based on a leading-order
truncation of the expansion at HERA: the small $x$ expansion 
breaks down for any reasonable value of $\alpha_s$ \brus.

To understand this result we will now consider analytically
the asymptotic behaviour of $P_{k}$ at large values of $\Xi$.
The asymptotic behaviour of the splitting functions 
may be determined through their definition as the inverse Mellin 
transform of the
anomalous dimension $\gamma$ \muel: at LLx
\eqn\loinvmel{P_0(\Xi)=\int_{-i\infty}^{i\infty}\!{dN\over2\pi
ia}\, e^{\xi N} \gamma_0(\bas/N)
=-\int_{-i\infty}^{i\infty}\!{d \gamma_0\over2\pi
i}\, e^{\xi \bas\chi_0(\gamma_0)}\gamma_0\chi_0'(\gamma_0) ,}
where we have used eq.~\expduallo\ to change integration variable 
$dN=\bas\chi_0'(\gamma_0) d\gamma_0$.
The asymptotic expansion of $P_0(\Xi)$ as $\xi\to\infty$ can then
be determined  by the saddle point method: a straightforward
computation, remembering that the only real minimum of
$\chi_0(\gamma_0)$ in the range $0<\gamma_0<1$ is at $\gamma_0=\half$
leads to 
\eqn\loasexp{P_0(\Xi)\twiddles{\Xi}\left(
\frac{\chi_0(\half)}{2\pi\chi_0''(\half)}\right)^{1/2}
\frac{\chi_0(\half)}{\Xi^{3/2}}
\,e^{\Xi} \left[1+\frac{\chi_0(\half)\chi_0''''(\half)}
{8({\chi''_0}(\half))^2}\frac{1}{\Xi}+O\left({1\over\Xi^2}\right)\right],}
where $\Xi$ is given by eq.~\xidef, while the $n$-th derivatives 
$\chi_0^{(n)}(\half)=4N_c\,n!(2^{n+1}-1)\zeta(n+1)$ when $n$ is even (all
odd derivatives vanish).

The asymptotic behaviour of the next-to-leading correction $P_1$ can be
determined in a similar way:
\eqn\nloinvmel{\eqalign{P_1(\Xi)&
=-\int_{-i\infty}^{i\infty}\!{dN\over2\pi
ia}\, e^{\xi N}
{\chi_1[\gamma_0(\bas/N)]\over\chi_0'[\gamma_0(\bas/N)]}
=-\int_{-i\infty}^{i\infty}\!{d\gamma_0\over2\pi
i}e^{\xi \bas\chi_0(\gamma_0)}\chi_1(\gamma_0)\cr
&\twiddles{\Xi}
\left(\frac{\chi_0(\half)}{2\pi\chi_0''(\half)}\right)^{1/2}
\frac{\chi_1(\half)}{\Xi^{1/2}}e^{\Xi}
\left[1+O\left({1\over\Xi}\right)\right],\cr
}}
where in the last step we have assumed $\chi_1(M)$ to be regular at
$M=\half$. This requirement can always be achieved by choice of
factorization scheme: in particular
$\chi_1^{FL}(\half)=-71.64N_c^2 -0.52N_cn_f - 10.7 n_f/N_c$ 
is finite~\fl\foot{Note that the results given in both 
\fl\ and \fad\ are numerically incorrect.} and in the \MS\ 
factorization scheme 
$\chi_1(\half)$ is effectively (see eq.\chimsb) $\chi_1^{FL}(\half)+
b_0 N_c[(4\ln 2)^2-\smallfrac{2\pi^2}{3}]$.  

It follows that asymptotically at large $\Xi$ in the \MS\ scheme 
\eqn\asrat{{P_1(\Xi)\over
P_0(\Xi)}\twiddles{\Xi} {\chi_1(\half)\over\chi_0(\half)}\Xi +O(1):}
the next-to-leading correction, despite being suppressed by a factor
of $\alpha_s$, rises linearly with $\Xi$ and hence with $\ln 1/x$ 
at small $x$, and becomes eventually dominant. In terms of the 
anomalous dimension $\gamma$, this means that the ratio 
$A^{(n)}_1/A^{(n)}_0$ of the NLLx coefficients to the LLx coefficients
in the expansions \expdef\ rises linearly with $n$. It is clear that 
the origin of the rise  
is the simple pole of $\gamma_1$ eq.~\expdualnlo, viewed as a
function of $\gamma_0$, at $\gamma_0=\half$.
Because the denominator of eq.~\expdualnlo\ vanishes linearly at
$\gamma_0$, this  simple pole is present whenever $\chi_1(\half)$ 
has a finite nonzero value. 

In factorization schemes where  $\chi_1(M)$ diverges at $M=1/2$ 
 the singularity in $\gamma_1$ will be stronger, and so 
$P_1$ will rise more rapidly at large $\xi$. Such factorization 
schemes have been
considered in the literature as being possibly more appropriate at
small $x$: in particular, the $Q_0$--schemes~\ciafqz, the SDIS
scheme~\sdis\ and the ``physical'' scheme (GDIS)~\refs{\phys,\mom}, 
all of which have the advantage of reducing the size of the leading
perturbative corrections in the quark sector.
This reduction is accomplished at the expense of introducing
a singularity in $\chi_1(M)$ at $M={1\over 2}$: for instance in the
$Q_0$--scheme 
(i.e. eq.\nlochi) $\chi_1(M)$ has a simple pole at $M=\half$ due 
to the vanishing of $\chi_0'(\half)$. 

It is easy to see  that if the NLO anomalous dimension has a 
simple pole at the location of the LO saddle, this dominates the 
asymptotic behaviour of integral~\nloinvmel. In this case, 
on top of the contribution of
eq.~\nloinvmel, there is a  further contribution from the 
residue of the pole:
\eqn\nloinvmelsing{P_1^{\rm sing.}(\Xi)\twiddles{\Xi}-\half
{\rm Res}[ \chi_1]e^{\Xi}+\ldots,}
where ${\rm Res}[ \chi_1]$ is the residue of the simple pole of
$\chi_1(M)$ at $M=\half$: it is equal to $\half b_0\chi_0(\half)$ in
all of the singular schemes mentioned above. The ratio $P_1/P_0$ now 
rises as\foot{A similar result  
was found in ref.\muel.} $\Xi^{3/2}$:
\eqn\asrats{
{P_1(\Xi)\over
P_0(\Xi)}\twiddles{\Xi} 
\left({\chi_1^{FL}(\half)\over\chi_0(\half)}+k\right)\Xi
-b_0\left(\frac{\pi\chi_0''(\half)}
{8\chi_0(\half)}\right)^{1/2} \Xi^{3/2}
\left[1-\frac{\chi_0(\half)\chi_0''''(\half)}
{8({\chi''_0}(\half))^2}\frac{1}{\Xi}\right]+O(1),}
where $k=2b_0,-2b_0,0$ in the $Q_0$-DIS \ciafqz, SDIS \sdis\ and 
GDIS \refs{\phys,\mom} schemes respectively. 
In all such schemes, the small $x$ expansion appears thus to be particularly
badly behaved. 

\topinsert
\vskip-2.5truecm
\vbox{
\hfil\epsfxsize=8.5truecm\epsfbox{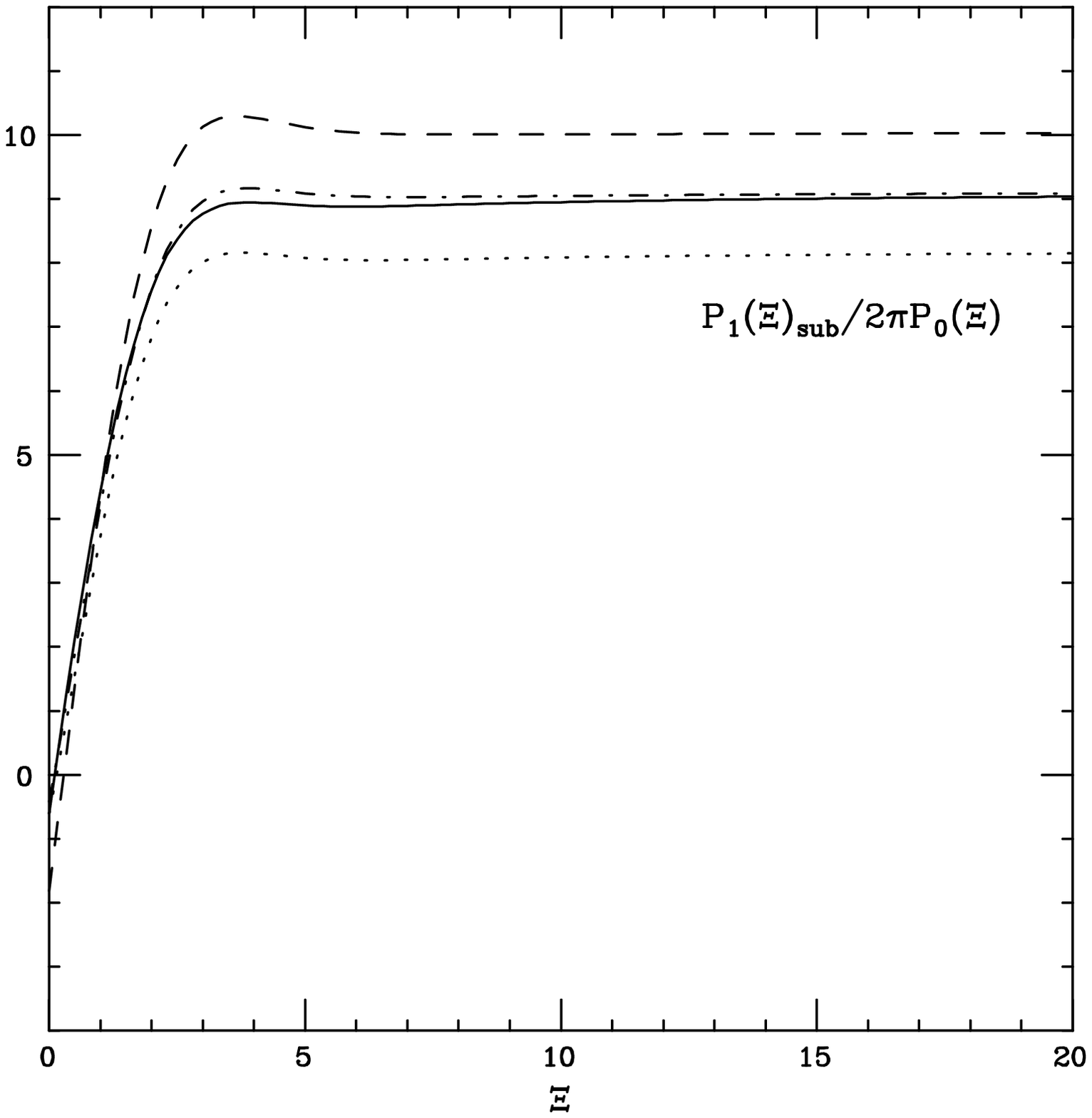}
\vskip-2.2truecm
\bigskip
\hskip4truecm\hbox{\vbox{\footnotefont\baselineskip6pt\narrower\noindent
Figure 4: As fig.~3, but with the asymptotic behaviours \asrat\ and
\asrats\ calculated analytically subtracted from the previous results 
computed numerically. Note the change in scale on the vertical axis. 
}}\hskip4truecm}
\medskip
\vskip-0.3truecm
\endinsert

In Fig.~4 the splitting function ratio $P_1/P_0$ is plotted again 
but now with the asymptotic results \asrat\ and \asrats\ subtracted.
It is clear that the asymptotic behaviour sets in surprisingly quickly:
already at $\Xi\gsim 3$ the subtracted ratio becomes
constant. It follows that what is left of $P_1$ after the subtraction is
no longer unnaturally large: the asymptotic growth of the ratios 
\asrat\ and \asrats\ is entirely responsible for the breakdown of
small $x$ perturbation theory at NLLx. This could have been 
anticipated even before the calculation of ~\fl: the
NLLx correction to the LLx splitting function will inevitably 
become large asymptotically unless the NLLx correction 
to $\chi(\half)$ vanishes.

Pursuing the argument to higher orders, it is apparent that the
anomalous dimensions $\gamma_i$ have higher order poles as the order
of the expansion increases: $\gamma_2$ eq.~\expdualnnlo\ 
has a triple pole, and in general $\gamma_n$ has a $(2n-1)$--th
order pole. Consequently, the associated splitting functions 
will display stronger and stronger rises with $\xi$. 
It follows that the small $x$ expansion eqns.~\lsexp,\lsexpsf\ 
inevitably breaks down at small $x$: as $\Xi$ grows, the
higher orders of the expansion become more and more important.
This means that the leading contributions at small $x$ have not been
properly resummed.

The origin of this failure can be simply understood by
recalling that at fixed coupling the leading asymptotic 
small $x$ behaviour of the solution to the $\xi$-evolution 
eq.~\lrge\ is given by $x^{\bas\chi(M_s)}$
where  $M_s$ is the position of the saddle point in $M$
(so at leading order $\chi=\chi_0$ and $M_s=\half$). 
When solving the small $x$ Altarelli-Parisi equation eq.~\trge,
this growth at small $x$,
rather than being generated by solving an evolution equation in $\xi$
eq.~\lrge\ (with $\xi$-independent anomalous dimensions), 
is included in the splitting functions. But
expanding the corrections to the LLx asymptotic behaviour in 
powers of $\alpha_s$
\eqn\nlsxappr{
x^{\bas
\left(\chi_0(M_s)+\as\chi_1(M_s)+\dots\right)}
=x^{\bas\chi_0(\half)}\left[1+\as \Xi
{\chi_1(\half)\over\chi_0(\half)}+\ldots\right],}
it is clear that the LLx asymptotic behaviour can be modified by
subleading terms only if these rise with $\Xi$. For
instance, the NLLx correction can only be
generated if $P_1/P_0$ rises linearly with $\Xi$, with slope  
$\chi_1(\half)/\chi_0(\half)$, as indeed we found above in 
eq.~\asrat. At higher orders in $\alpha_s$ the LLx behaviour
receives corrections proportional to higher powers of $\Xi$, and
correspondingly the higher order $P_n$  have higher order
poles. However these corrections are no longer given by a trivial
exponentiation of the NLLx result.

The bad behaviour of the small $x$ expansion is thus due to the fact that the
asymptotic behaviour of the solution to the evolution equations 
at large $\Xi$ does not coincide with the LLx prediction.   The
subsequent mismatch in the order of subleading corrections  makes
a nonsense of the perturbative expansion \expdef\ and \expdefsf. This 
can only be corrected by suitably reorganizing the expansion \chiexp\ 
in order to properly resum the large corrections. 

Since a change in the factorization scheme mixes different orders 
in the perturbative expansion,  different scheme choices may 
be thought of as resummations. Furthermore 
just as there are scheme choices which make 
the perturbative expansion less stable, so there are choices 
which can improve it. More precisely, we can resum the large
corrections by choosing the scheme in such a way that the 
anomalous dimensions 
eqns.~\expdualnlo,\expdualnnlo,... are all regular at 
$\gamma_0=\half$ order by order. This is always possible because if all
$\gamma_i$ with $i\le i_0-1$ are regular at $\gamma_0=\half$, then 
$\gamma_{i_0}$ has a simple pole at $\gamma_0=\half$ (assuming that all
$\chi_i(M)$ are regular there), which can be removed by choosing a
scheme which subtracts a constant from $\chi_i(M)$ equal to 
the numerators of eqns.~\expdualnlo,\expdualnnlo\ and
their  higher order generalizations. Necessary conditions for 
a satisfactory perturbative scheme are thus that in the new scheme
\eqn\cons{
\chi_1(\half)=0, \qquad \chi_2(\half) 
=\half(\chi_{1}'(\half))^2/\chi_{0}''(\half),\qquad\cdots.}

Clearly these conditions are not very restrictive. In fact, 
since there is only one condition at each perturbative order, 
they can be imposed simply by a choice of renormalization scale, 
\ie\  by the replacement of $\alpha_s(Q^2)$ with $\alpha_s((kQ)^2)$, 
where $k$ may itself be expanded as a series in $\alpha_s$. 
Then for example at NLLx 
$A_1^{(n)}\to A_1^{(n)} +A_0^{(n)}nb_0\log k^2$, and the linear rise
of  $A_1^{(n)}/A_0^{(n)}$ in \MS\ factorization schemes 
may be eliminated by choosing 
$b_0\log k=-\half\chi_1(\half)/\chi_0(\half)$,
which gives $k\simeq 300$. Choosing such a large scale does indeed
lead to stable perturbative behaviour (see fig.~5). However it is also
clearly a fine tuning: varying the scale by a factor of two either
side leads to huge variations in the relative size of the perturbative
correction. Other scale choices, such as BLM, designed to reduce the size of 
subleading corrections have been considered in ref.\blm.\foot{It 
is also possible to remove the singularity \nloinvmelsing\ in 
singular schemes such as Q$_0$-DIS by tuning the scale. However 
now the choice of scale depends on $x$: $k\sim
k_1\exp (k_2\sqrt\xi+k_3/\sqrt\xi)$, where each of $k_1$, $k_2$ and
$k_3$ all require fine tuning if the perturbative 
expansion is to be stable. Such a scale might also be justified 
through a BLM prescription~\rst.} At NLLx it is even possible 
to fine tune the choice of scheme such that
$\chi_1(M)\equiv 0$, by choosing the gluon normalization 
factor $u(M)$ as a solution to the first order differential equation
$(\ln u)'=\chi_1(M)/b_0\chi_0(M)$ (cf. \schchi): then
$\gamma_1(\alpha_s/N)\equiv 0$ and all NLLx corrections 
have been removed from the large eigenvalue of evolution.

The physical meaning of these scheme choices is that, after
the scheme change, the large corrections to the LLx asymptotic small $x$
behaviour are absorbed into the $x$-dependent
initial condition to the perturbative evolution in $Q^2$.
However because these large corrections have a large scheme
dependence, it seems pointless to consider them at any finite order:
all contributions to the asymptotic behaviour
should be resummed to all orders, and then included in the LLx
anomalous dimension. To this purpose it is sufficient to subtract 
$Q^2$-independent contributions in eq.~\chiexp: 
\eqn\reshexp{
\chi(M;\as)=c(\as)+\tilde\chi_0(M)+\as\tilde\chi_1(M)
+\dots,}
where $\tilde\chi_i(M)\equiv\chi_i(M)-c_i$, $i=0,1,2,\ldots$ and 
formally $c(\as)\equiv c_0+\sum_{n=1}^\infty c_n \as^n$. 
The subtractions
$c_1,c_2,\ldots$ are then  fixed by the criteria \cons: we need 
\eqn\subtns{c_1=\chi_1(\half),\qquad c_2=\chi_2(\half) 
-\half(\chi_{1}'(\half))^2/\chi_{0}''(\half),\qquad\cdots.}
Since the resummed splitting function will be independent of $c_0$,
there is in principle no need to fix its value; however it is
convenient to choose  $c_0=\chi_0(1/2)$ as with this choice
the parameter
\eqn\lamdef{\lambda= \bas c(\as)}
has a direct physical meaning.

We can now use the expansion \reshexp\ of $\chi$ in eq.\dual\ to
determine $\tilde\gamma_i$ order by order, treating $c(\as)$ as
leading order. At LLx the `resummed' anomalous dimension 
$\tilde \gamma_0(N;\as)$ is then the solution of
\eqn\resgam{
c(\as)+\tilde\chi_0[\tilde\gamma_0]=N/\bas.}
This implies that 
$\tilde \gamma_0(N;\as)=
\gamma_0[\bas/(N-(\lambda-\bas c_0))]$ 
and consequently that the LLX splitting 
function 
\eqn\rescz{\tilde P_0(x;\as)=P_0(\Xi)e^{(\lambda- \bas c_0)\xi}.}
In particular the asymptotic behaviour \loasexp\ becomes 
\eqn\asres{
\tilde P_0(x;\as)\twiddles{\Xi}\left(
\frac{\chi_0(\half)}{2\pi\chi_0''(\half)}\right)^{1/2}
\frac{\chi_0(\half)}{\Xi^{3/2}}\,e^{\lambda\xi}
\left[1+O\left({1\over\Xi}\right)\right].}
The parameter $\lambda$ thus determines the nature
of the asymptotic small
$x$ behaviour. The expansion in powers of $\alpha_s$ is now 
well-behaved: due to the conditions \subtns\ all higher order 
$P_i(x)$ behave in the same way as $P_0$ at large $\xi$, 
and thus all the corrections to $P_0$ are down by powers of 
$\alpha_s$ uniformly in $x$. The ratio $\tilde P_1/\tilde P_0$ is
shown in fig.~5: it is indeed uniformly bounded and not too unreasonably 
large.\foot{In fact it is identical to the subtracted ratio
in DIS plotted in fig.~4: this is because 
$${\frac{\chi_1(\half)}{\chi_0(\half)}\Xi P_0(\Xi)=
-\chi_1(\half)\int_{-i\infty}^{i\infty}\frac{d\gamma_0}{2\pi i}
\gamma_0\frac{\partial}{\partial\gamma_0}e^{\bas\xi\chi_0(\gamma_0)}
=c_1\int_{-i\infty}^{i\infty}\frac{dN}{2\pi ia} 
\frac{e^{N\xi}}{(-\chi_0'(\gamma_0(\bas/N)))},}$$
by an integration by parts and change of variables.}

\topinsert
\vskip-2.5truecm
\vbox{
\hfil\epsfxsize=8.5truecm\epsfbox{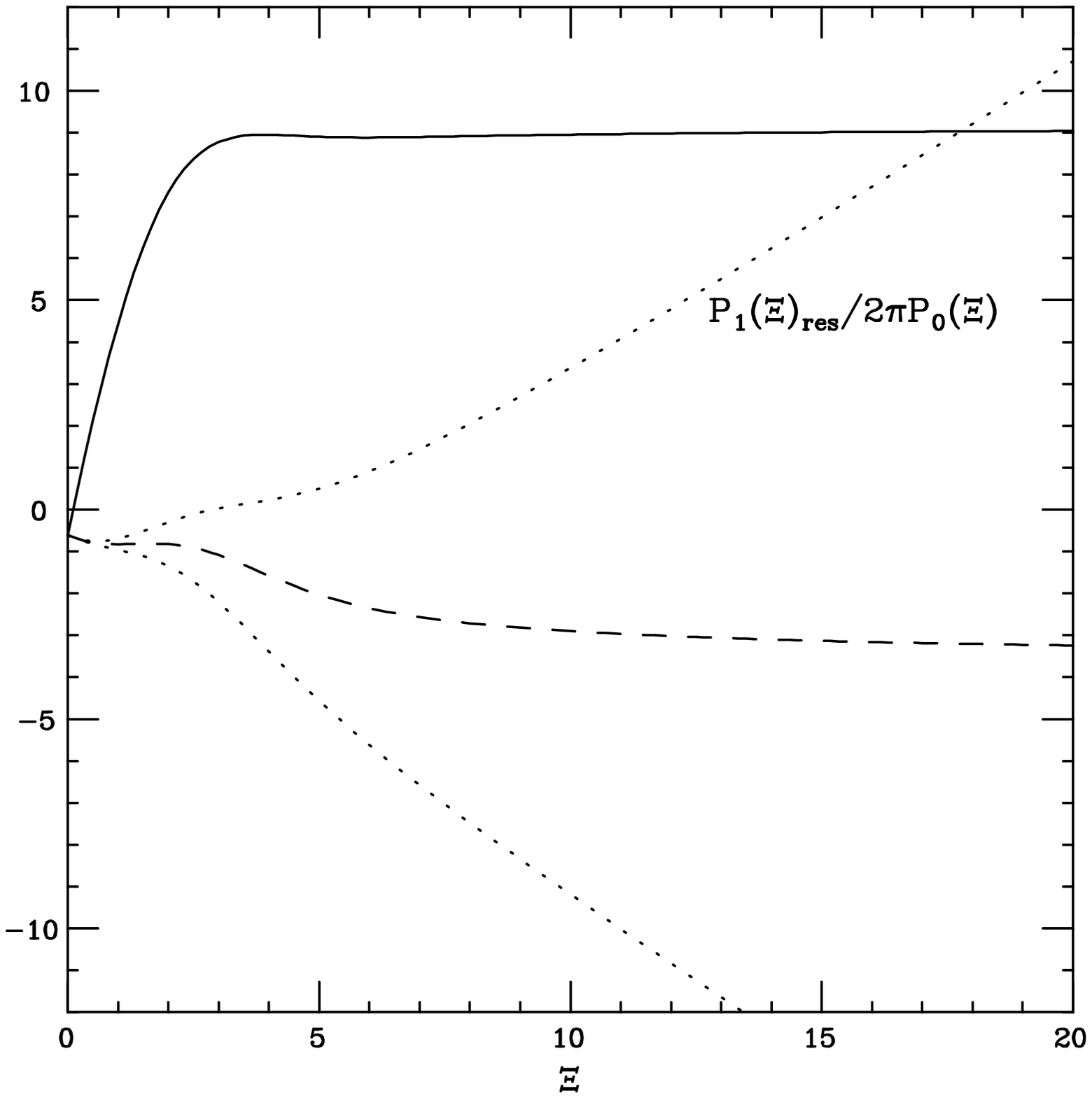}
\vskip-2.0truecm
\bigskip
\hskip4truecm\hbox{\vbox{\footnotefont\baselineskip6pt\narrower\noindent
Figure 5: As fig.~3, but after various resummations: the subtraction 
\reshexp, the fine tuned scale (dashed), and scales a factor
of two either side of it (dotted).  
}}\hskip4truecm}
\medskip
\endinsert
The reorganization of the perturbative expansion
\reshexp\ can be viewed as an effective resummation of the higher
orders of the expansion. Since $c(\as)-c_0$ is of $O(\as)$,
eq.~\rescz\ implies that  $\tilde P_0$ and $P_0$ differ by
a series of formally subleading contributions. However, the parameter
$\lambda$~\lamdef\ which summarizes the asymptotic behaviour at small
$x$ is treated as $\alpha_s$-independent, and thus
included in the leading order eq.\resgam. This `order transmutation'  
effectively  resums into the LLx anomalous dimension the
all-order behaviour as given by $\lambda$. 
However, to determine the value of $\lambda$ it may be 
necessary to use arguments
which go beyond mass factorised perturbation theory. In particular, the
unitarity constraint would suggest that $\lambda\leq 0$: if
$\lambda$ were positive the resulting powerlike growth in the
splitting function at small $x$ would drive a corresponding rise in
the cross-section, which would ultimately violate the Froissart bound. 

The removal of the unbounded growth of formally subleading corrections
at small $x$, achieved by the resummation described
above, is a necessary prerequisite for a consistent small $x$
resummation of Altarelli-Parisi evolution. Although our resummation
does not resolve the instability in the small $x$ evolution equation
discussed in \refs{\ross-\ABB}, it does show that this is 
a separate issue. Indeed, the instability is clearly related to the 
shape of $\chi$ as $M\to 0$ and $M\to 1$, whereas the resummation 
criteria \cons\ refer to $M=\half$. At small $M$ (and thus large
$Q^2$), the relevant approximation is to use the conventional 
Altarelli-Parisi equation: from this it may readily be inferred (using a
duality argument \afp\ inverse to that used to obtain \dual) that 
the resummed kernel must always be finite and positive at $M=0$, which is
probably sufficient to cure the instability. Possibly related 
attempts to deal with these instabilities have been presented 
in ref.\refs{\salami}. 

To conclude, we have shown that the poor behaviour of the small $x$ expansion
which is manifested~\refs{\brus,\blum} in the NLLx splitting functions
computed from the recent Fadin-Lipatov determination~\fl\ of 
the next-to-leading high energy QCD asymptotics can be traced to the
fact that the formally NLLx corrections to the LLx contributions to
the splitting functions are not truly subleading at small $x$. We 
have shown that this problem persists to all orders, and is 
related to the fact that the leading small $x$ behaviour is 
not given by the leading order term of the small $x$ 
expansion, but rather must come from an all-order resummation. 
We have demonstrated that a reorganization of the perturbative 
expansion is necessary, and given criteria eq.\cons\  which
must be met if such a resummation is to be successful. We further 
constructed a resummation which meets these criteria, but depends 
on a new parameter $\lambda$ eq.\lamdef\
which controls the asymptotic growth at small $x$. 
A complete resummation of Altarelli-Parisi splitting functions at 
small $x$ might now be achieved through 
careful matching in the high $Q^2$ region.
 
\bigskip
{\bf Acknowledgements}: We thank G.~Altarelli for several discussions
and a critical reading of the manuscript.
This work was supported in part by a PPARC Visiting Fellowship, and 
EU TMR  contract FMRX-CT98-0194 (DG 12 - MIHT). 

\vfill\eject
\listrefs
\vfill\eject
\bye